\documentclass[pra,showpacs,amsmath,amssymb]{revtex4}

\usepackage{graphicx}
\usepackage{dcolumn}
\usepackage{bm}

\begin{document}

\def\bra{\langle}  \def\ket{\rangle}
\def\ketpsi{| \psi \rangle}


\title{Macroscopic Entanglement in Quantum Computation}
\author{Akihisa Ukena}
\email{ukena@ASone.c.u-tokyo.ac.jp}
\affiliation{
Department of Basic Science, University of Tokyo, 
3-8-1 Komaba, Tokyo 153-8902, Japan
}
\affiliation{
PRESTO, Japan Science and Technology Corporation,
4-1-8 Honcho, Kawaguchi, Saitama, Japan
}
\author{Akira Shimizu}
\email{shmz@ASone.c.u-tokyo.ac.jp}
\affiliation{
Department of Basic Science, University of Tokyo, 
3-8-1 Komaba, Tokyo 153-8902, Japan
}
\affiliation{
PRESTO, Japan Science and Technology Corporation,
4-1-8 Honcho, Kawaguchi, Saitama, Japan
}
\date{\today}

\begin{abstract}
We investigate macroscopic entanglement of quantum states in quantum 
computers, where we say a quantum state is entangled macroscopically 
if the state has superposition of macroscopically distinct states.  
The index $p$ of the macroscopic entanglement is calculated as a 
function of the step of the computation, for Grover's quantum search 
algorithm and Shor's factoring algorithm.  It is found that whether 
macroscopically entangled states are used or not depends on the 
numbers and properties of the solutions to the problem to be solved.  
When the solutions are such that the problem becomes hard in the 
sense that classical algorithms take more than polynomial steps to 
find a solution, macroscopically entangled states are always used 
in Grover's algorithm and almost always used in Shor's algorithm.  
Since they are representative algorithms for unstructured and 
structured problems, respectively, our results support strongly 
the conjecture that quantum computers utilize macroscopically 
entangled states when they solve hard problems much faster than any 
classical algorithms.
\end{abstract}


\maketitle

\section{Introduction}\label{sec:intro}

Quantum computers are many-body systems with a large but finite number $L$
of qubits because the efficiency of computation becomes
relevant only when $L$ is large \cite{NC}.
Although the entanglement is considered to play crucial 
roles in speedup as compared with classical computers \cite{Vidal,JL,US04},
there are many types of entanglement 
and many measures or indices of entanglement
for many-body systems 
\cite{Bennett,miyake,SM02,Meyer_Wallach,Stockton,Syljuasen,Verstraete,
MSS05,SS03}.
It is therefore interesting to explore which type of 
entanglement is relevant and which measure or index quantifies
such entanglement.

Recently, an index $p$ of {\em macroscopic entanglement}, 
which implies that 
the state has superposition of macroscopically distinct states,
was proposed 
and studied in Refs.~\cite{SM02,MSS05,SS03}:
A pure state, 
which is homogeneous or effectively homogeneous
(see Refs.~\cite{SM02,SS03,US04} and 
Secs.~\ref{sec:Grover-analytic}-\ref{sec:shor}),  
is entangled macroscopically iff it has $p=2$.
Although the term `macroscopic entanglement' are sometimes 
used in different ways, 
we follow these references.
Most importantly, 
the macroscopic entanglement {\em as defined by this index}
is directly related to fundamental stabilities 
of many-body quantum states 
(see Ref.~\cite{SM02} and Sec.~\ref{sec:p}).
Moreover, it was shown rigorously that 
any {\em pure} states with $p=2$ in systems of {\em finite} $L$ 
do {\em not} approach pure states in the {\em infinite} 
system as $L \to \infty$ \cite{SM02,Hspaces}.
Noting that macroscopically  entangled states
have such anomalous properties, 
one of the authors gave the conjecture that 
a quantum computer should utilize 
macroscopically  entangled states
in some stages of the computation
when it performs a quantum algorithm that is much more efficient than 
any classical algorithms 
(Ref.~\cite{conjecture} and Sec.~\ref{sec:conj}).

In the previous paper \cite{US04}, we confirmed this conjecture
in the case of Shor's factoring algorithm \cite{Shor_1,EJ}.
Since Shor's algorithm is a representative quantum algorithm
for solving structured problems \cite{NC}, 
it is very interesting to study whether
the conjecture is correct in the case of 
quantum algorithms for solving {\em unstructured} problems.
Furthermore, in Ref.~\cite{US04} we identified macroscopic entanglement
of only two states, the state after the modular exponentiation and
the final state, among many quantum states appearing in 
Shor's algorithm, because it was difficult to identify 
macroscopic entanglement of general states.
It is also interesting to study whether the other states
in Shor's algorithm are entangled macroscopically.
Recently, an efficient method
of calculating the index $p$ 
was developed and successfully applied to many-magnon states \cite{MSS05}
and to chaotic many-body states \cite{SS03}.
This method, called the variance-covariance matrix (VCM) method, 
enables us to study the macroscopic entanglement of general states.

In this paper, we study macroscopic entanglement
of quantum states of quantum computers performing 
Grover's quantum search algorithm \cite{Grover},
which is a representative algorithm
for solving unstructured problems \cite{NC}.
The results show that the above conjecture is correct also in this case.
We also investigate macroscopic entanglement
of all quantum states appearing in Shor's factoring algorithm,
using the VCM method,
and the above conjecture is confirmed again.
These results suggest the relevance of 
macroscopic entanglement in solving hard problems by 
quantum computers.

The paper is organized as follows.
In Sec.~\ref{sec:p}, we briefly explain the index $p$ of the 
macroscopic entanglement.
Section \ref{sec:conj} restates the conjecture 
in a more detailed form than 
Refs.~\cite{US04} and \cite{conjecture}.
Analytic results for Grover's quantum search algorithm
are given in Sec.~\ref{sec:Grover-analytic}, by which 
the conjecture is confirmed.
To see details of evolution of entanglement, 
we present results of 
numerical simulations of a quantum computer that performs 
Grover's quantum search algorithm
in Sec.~\ref{sec:Grover-numerical}.
Shor's factoring algorithm is studied 
in Sec.~\ref{sec:shor},
and the conjecture is again confirmed.
Discussions and conclusions are given in Sec.~\ref{sec:dc}.

\section{Index of macroscopic entanglement}\label{sec:p}

The index $p$ of macroscopic entanglement
was proposed and studies in Refs.~\cite{SM02,MSS05,SS03}.
In this section, we briefly explain it 
to establish the notations and for reader's convenience.


\subsection{Definition}

We consider quantum states which are
homogeneous, or effectively homogeneous as in 
quantum chaotic systems \cite{SS03} and in quantum computers
(see Secs.~\ref{sec:Grover-analytic}-\ref{sec:shor} and 
Ref.~\cite{US04}).
For such a state, 
we consider a family of similar states of various sizes, 
and denote the state of size $L$ by $\hat \rho(L)$.
We say $\hat \rho(L)$ (or a quantum system) is {\em macroscopic} if
for every quantity of interest
the term that is leading order in $L$ gives the dominant contribution
\cite{SM05}.
We say $\hat \rho(L)$ is {\em entangled macroscopically}
if the state has `superposition of macroscopically distinct states'
\cite{SM02,MSS05,SM05,custom}. 

In general, however, the term in the quotation marks 
is quite ambiguous \cite{MSS05,SM05}. 
For example, 
it would be obvious that 
a `cat state'
$
| \psi_{\rm cat}(L) \rangle 
\equiv
(
|00 \cdots 0 \rangle
+
|11 \cdots 1 \rangle
)/\sqrt{2}
$
of a quantum computer with $L$ qubits 
has such superposition,
whereas how about the following state \cite{SM05}?
\begin{equation}
| \psi_{\rm dws}(L) \rangle
\equiv
{1 \over \sqrt{L+1}}
\left(
|000 \cdots 0 \rangle
+
|100 \cdots 0 \rangle
+
|110 \cdots 0 \rangle
+
\cdots
+
|111 \cdots 1 \rangle
\right).
\label{eq:dws}\end{equation}

Recently, 
a reasonable index $p$ of macroscopic entanglement
was proposed for pure states \cite{SM02,MSS05}.
For $\hat \rho(L) = | \psi(L) \rangle \langle \psi(L) |$, 
this index is defined by \cite{O}
\begin{equation}
\max_{\hat A} \langle \psi(L) | 
\Delta \hat A^\dagger \Delta \hat A
| \psi(L) \rangle = \mathcal{O}(L^p),
\label{def:p}\end{equation}
where 
$\Delta \hat A \equiv \hat A 
- \langle \psi(L) | \hat A | \psi(L) \rangle$
and 
$\hat{A}$ is an additive operator, i.e., 
a sum of local operators \cite{SM02}; 
see, e.g., Eqs.~(\ref{add_op_0}) below.

The index $p$ 
ranges over 
$1 \leq p \leq 2$ (see Appendix \ref{appendix_range}).
According to Refs.~\cite{SM02,MSS05}, 
a pure state is entangled macroscopically if 
$p=2$ \cite{custom}, 
whereas it may be entangled but {\em not} macroscopically if $p<2$.
For example, 
the `W state,'
$
| \psi_{\rm W}(L) \rangle
\equiv
{1 \over \sqrt{L}}
\left(
|100 \cdots 0 \rangle
+
|010 \cdots 0 \rangle
+
\cdots
+
|000 \cdots 1 \rangle
\right)
$,
has $p=1$,
hence its entanglement is {\em not} macroscopic \cite{SM02,MSS05}.
This is reasonable because
the W state corresponds to 
a single-magnon state in a ferromagnet, 
which is quite a normal state.
Since both the W state and product states have $p=1$, 
they belong to the same class
with respect to the macroscopic entanglement.
On the other hand, we can show that 
$p=2$ for $|\psi_{\rm dws} \rangle$ of Eq.~(\ref{eq:dws}), 
hence is entangled macroscopically.
Physically, $|\psi_{\rm dws} \rangle$ is superposition of 
single-domain-wall states which have different positions of 
the domain wall.

It is worth mentioning that
a generalized index $q$, which can be 
applied to mixed states as well, 
and a method for detecting macroscopic entanglement
were proposed in Ref.~\cite{SM05}.
However, we do not discuss these points in this paper
because 
we treat pure states only.

\subsection{Efficient method of calculating $p$}

Evaluation of a measure or index of entanglement often becomes intractable
for large $L$.
Fortunately, this is not the case for $p$, i.e., 
there is an efficient method of calculating $p$ 
\cite{MSS05,SS03}.

For qubit systems, an additive operator can be expressed as
\begin{equation}
\hat{A}=\sum_{l=1}^{L} 
\left[
c_{l 0} \hat 1(l)+
\sum_{\alpha=x,y,z} 
c_{l \alpha} \, \hat{\sigma}_{\alpha}(l)
\right].
\label{add_op_0}
\end{equation} 
Here, 
$\hat 1(l)$  and
$\hat{\sigma}_\alpha(l)$ ($\alpha=x,y,z$) denote 
the identity and 
the Pauli operators, respectively, acting on the qubit at site $l$,
and $c_{l 0}$'s and $c_{l \alpha}$'s are complex numbers which 
are 
independent of $L$.
For convenience, we here allow non-hermitian additive operators,
as in Refs.~\cite{SM02,MSS05}.
Since we are interested in $\Delta \hat A$, 
we henceforth drop 
$\sum_l c_{l 0} \hat 1(l)$
from Eq.~(\ref{add_op_0}) \cite{MSS05}, i.e., 
\begin{equation}
\hat{A}=\sum_{l=1}^{L} 
\sum_{\alpha=x,y,z} 
c_{l \alpha} \, \hat{\sigma}_{\alpha}(l).
\label{add_op}
\end{equation} 
Without loss of generality, we normalize $c_{l \alpha}$'s as
\begin{equation}
\sum_{l=1}^{L}\sum_{\alpha=x,y,z}|c_{l \alpha}|^{2}
=L.
\label{eq:normalization}\end{equation}
The $\max_{\hat A}$ in Eq.~(\ref{def:p}) is thus taken 
over all possible choices of $c_{l \alpha}$'s satisfying this 
condition.

Let us define the variance-covariance matrix (VCM) for 
a pure state $| \psi(L) \rangle$ by
\begin{equation}
V_{\alpha l \beta l'}=\langle \psi(L) |
\Delta \hat{\sigma}_{\alpha}(l) \Delta \hat{\sigma}_{\beta}(l') 
|\psi(L) \rangle,
\label{VCM}\end{equation}
where 
$\Delta \hat{\sigma}_{\alpha}(l) 
\equiv
\hat{\sigma}_{\alpha}(l) - 
\langle \psi(L) | \hat{\sigma}_{\alpha}(l) |\psi(L) \rangle$, 
$\alpha, \beta = x, y, z$, and
$l, l' = 1, 2, \cdots, L$.
It was shown in Ref.~\cite{MSS05} that 
the maximum eigenvalue $e_{\rm max}(L)$ of the VCM is related to $p$ 
of $| \psi(L) \rangle$ as
\begin{equation}
e_{\rm max}(L) = \mathcal{O}(L^{p-1}).
\label{emax=Lp-1}\end{equation}
Hence, we can calculate $p$ by calculating $e_{\rm max}(L)$, which 
can be done in a Poly$(L)$ time because the VCM is a
$3L \times 3L$ hermitian matrix.
Furthermore, 
from the eigenvector(s) 
of the VCM 
corresponding to 
$e_{\rm max}(L)$, 
we can find 
the {\em maximally-fluctuating additive operator(s)}
$\hat{A}_{\rm max}$, which satisfies \cite{MSS05,MSunpub}
\begin{equation}
\max_{\hat A} \langle \psi(L) | \Delta \hat A^\dagger \Delta \hat A
| \psi(L) \rangle
=
\langle \psi(L) | 
\Delta \hat{A}_{\rm max}^\dagger \Delta \hat{A}_{\rm max} 
|\psi(L) \rangle
\sim
e_{\rm max}(L)L
=\mathcal{O}(L^p).
\label{e_max}
\end{equation}

\subsection{Physical properties of $p$}


It was shown in Ref.~\cite{SM02} that 
$p$ is directly related to fundamental stabilities, 
including decoherence and stability against local measurements, 
of many-body states \cite{CPandp}.
Regarding decoherence by weak perturbations from 
noises or environments, 
it was shown that for any state 
with $p=1$ its decoherence rate $\Gamma$ by {\em any} of such perturbations
{\em never} exceeds $\mathcal{O}(L)$.
For a state with $p=2$, on the other hand, 
it is possible to {\em theoretically construct} a noise or environment 
that makes $\Gamma$ of the state $\mathcal{O}(L^2)$.
This does {\em not} necessarily mean that 
such a fatal noise or environment {\em exists in real systems}:
It depends on physical situations \cite{SM02}.
In this view, 
a more fundamental stability is the 
stability against local measurements, which was proposed and 
defined in Ref.~\cite{SM02}.
From the theorem proved there, 
we can see that a state with $p=2$ is {\em not} stable
against local measurements, i.e., 
there exists a {\em local} observable by measurement of which 
the state changes drastically.
Furthermore, 
it was also proved rigorously that 
any {\em pure} states with $p=2$ in systems of {\em finite} $L$ 
do {\em not} approach pure states in the {\em infinite} 
system as $L \to \infty$ \cite{SM02,Hspaces}.

These observations indicate that pure states with $p=2$ 
are quite anomalous many-body states.
This led to the conjecture of Ref.~\cite{conjecture},
which will be restated more clearly in Sec.\ref{sec:conj}.
If this conjecture is correct, 
$p$ is one of indices that are most directly related to 
efficiency of quantum computation,
among many measures and indices of entanglement.
This conjecture has been confirmed in Ref.~\cite{US04}
for Shor's factoring algorithm, 
which is a representative algorithm for structured problems, and 
will be confirmed in this work
for Grover's quantum search algorithm, 
which is a representative algorithm for unstructured problems.



The index $p$ was also studied for 
many-magnon states in Ref.~\cite{MSS05},
and for many-body chaotic states in Ref.~\cite{SS03}.
Comparison with another measure of entanglement were also made 
in these references.
Most importantly, 
many states were found such that they are almost maximally 
entangled in another measure but their $p$ is minimum, $p=1$.
Many other states are also found 
such that their $p$ is maximum, $p=2$, but
their entanglement is small in another measure.
This demonstrates the well-known fact that 
a simple statement such as 
`more efficiency of computation requires more entanglement'
is meaningless unless the measure or index of
the entanglement is specified.
In this paper, we focus on the macroscopic entanglement
as measured by the index $p$.


\section{Conjecture on quantum computation}\label{sec:conj}

Consider a computational problem, 
the {\em fastest} classical algorithm for which takes
$Q_{\rm cl}^{\rm min}(L)$ steps 
when the size of the input is $L$ bits.
We consider quantum computers
that solves the same problem.
Here, to be definite, we mean quantum circuits by `quantum computers.'
There are many different circuits for a single 
quantum algorithm.
We say 
{\em a quantum algorithm $\mathcal{A}$ solves this problem
much faster than classical algorithms}
if some quantum circuit of 
$\mathcal{A}$ takes only $Q(L)$ steps
which is much less than $Q_{\rm cl}^{\rm min}(L)$
in the sense that
\begin{equation}
{\ln \left[\ln Q_{\rm cl}^{\rm min}(L) - \ln Q(L) \right]
\over
\ln L}
\geq \mbox{a positive constant, for large $L$}.
\label{eq:faster}\end{equation}
Here, in the case of the search problem of Grover, 
we count each oracle call as one step in defining 
$Q_{\rm cl}^{\rm min}(L)$ and $Q(L)$.
Let $\mathcal{L}_0(L)$ be the {\em minimum} number of qubits 
of such circuits that satisfy this inequality.
Our conjecture is as follows: 
If a quantum computer which has $\mathcal{L}(L)$ 
qubits performs $\mathcal{A}$ with $Q(L)$ steps that satisfies 
inequality (\ref{eq:faster}), 
and if 
\begin{equation}
\mathcal{L}(L) = \mathcal{L}_0(L) + O(1),
\label{eq:L=L0+O1}\end{equation}
where $O(1)$ denotes a constant that is independent of $L$,
then a macroscopically entangled state(s) 
of size $\mathcal{O}(L)$ or more 
appears in the quantum computer
during the computation \cite{O}. 
Although it would be hard to determine $\mathcal{L}_0(L)$
of a general algorithm, we assume 
optimistically in this paper that 
the simple circuits 
of Secs.~\ref{sec:Grover-numerical} and \ref{sec:shor}
for 
Grover's quantum search and Shor's factoring 
algorithms, respectively, 
would satisfy Eq.~(\ref{eq:L=L0+O1}).

In the case of Grover's quantum search algorithm, for example, 
the left-hand side of inequality (\ref{eq:faster}) approaches $1$ 
as $L \to \infty$ (thus the inequality is satisfied) 
if the number of the solutions $M$ is $\mathcal{O}(1)$.
Furthermore, it seems reasonable to assume that 
\begin{equation}
\mathcal{L}_0(L) \geq L
\label{L0gtrL}\end{equation}
because the index register requires $L$ qubits, 
and more qubits may be 
required for the oracle, the conditional phase shift, and so on.
As we will show in this paper, 
macroscopically entangled states of size $\mathcal{O}(L)$
indeed appear
during the computation if $M=\mathcal{O}(1)$.

Note that 
we are considering quantum computers which 
have almost the minimum number $\mathcal{L}(L)$ 
of qubits to perform $\mathcal{A}$.
This is because it is generally possible to mask 
macroscopic entanglement if one can add arbitrarily many qubits
to the computer.
We exclude such uninteresting possibility by imposing 
Eq.~(\ref{eq:L=L0+O1}). 
However, it is worth mentioning that one can easily 
generalize the above conjecture to the case where
a quantum error correction \cite{ftc} is used in the computer.
That is, although the number of {\em physical} qubits 
does not satisfy Eq.~(\ref{eq:L=L0+O1}) in such a case, 
one can identify each {\em logical} qubit with a large qubit.
Then, Eq.~(\ref{eq:L=L0+O1}) is satisfied for the number of the 
large qubits, and the conjecture 
can be applied if the macroscopic entanglement is defined 
in terms of the large qubits.

Since we only consider pure states in this paper, 
a macroscopically entangled state is a state with $p=2$.
Hence, the above conjecture may be stated roughly as follows:
{\em A quantum algorithm that is much faster than classical 
algorithms utilizes some state(s) with $p=2$}.
Note that this does {\em not} claim that 
{\em all} states with $p=2$ would be useful to fast 
quantum computation.
In particular, 
the conjecture does {\em not} claim that 
{\em all} macroscopically entangled states appearing in 
the computation would be relevant to fast quantum computation.
This point will be discussed in Secs.~\ref{sec:relevance-G}
and \ref{sec:relevance-S} by showing examples.

Furthermore, the following fact is worth mentioning.
For some macroscopically entangled states of size $L$, 
there are quantum circuits that 
convert a product state into 
such a state in $\mathcal{O}(L)$ 
steps {\em if the target state with $p=2$ is known} when the circuits are designed.
This fact has {\em nothing} to do with our conjecture.
The point is that 
in quantum computation 
the states that appear 
at, e.g., the middle point of the computation
are {\em unknown} when the circuits are designed,
because the states depend on the solutions of the problems to be solved.
The above conjecture claims the appearance of 
macroscopically entangled states in this case.



\section{Analytic results for Grover's quantum search algorithm}
\label{sec:Grover-analytic}

Consider the problem of finding a solution to  the equation $f(x) = 1$
among $N=2^L$ possibilities, where $f(x)$ is a function, 
$
f:\{0,1\}^{L} \mapsto \{0,1\}.
$
If the number $M$ of solutions is $\mathcal{O}(1)$,
this `search problem' is hard in the sense that 
classical algorithms take $\mathcal{O}(N)$ 
steps.
A quantum computer using 
Grover's quantum search algorithm solves this problem much faster,
with $\mathcal{O}(\sqrt{N})$ 
steps \cite{Grover,NC}, and 
inequality (\ref{eq:faster}) is satisfied.
Hence, Grover's  algorithm is a representative 
quantum algorithm for unstructured problems \cite{NC}.
We investigate the index $p$ of quantum states 
appearing in a quantum computer that performs this algorithm.
As a quantum computer, we consider the index register of $L$ qubits, 
and thus Eq.~(\ref{eq:L=L0+O1}) is clearly satisfied.

Suppose that the equation $f(x) = 1$ has $M$ solutions $x_1, \cdots, x_M$.
If we put
\begin{eqnarray}
| \alpha \rangle &\equiv& 
{1 \over \sqrt{N-M}} \sum_{x \neq x_1, \cdots, x_M} | x \rangle,
\label{eq:alpha}\\
| \beta \rangle &\equiv& 
{1 \over \sqrt{M}} \sum_{x = x_1, \cdots, x_M} | x \rangle,
\label{eq:beta}\end{eqnarray}
then the state $| \psi_0 \rangle$ just after the first Hadamard transformation 
(see Eq.~(\ref{Hadamard}) below) is represented as \cite{NC}
\begin{equation}
| \psi_0 \rangle = 
\cos {\theta \over 2} \ | \alpha \rangle 
+ 
\sin {\theta \over 2} \ | \beta \rangle.
\label{eq:psi0=}\end{equation}
Here, 
the angle $\theta$ is given by
\begin{equation}
\cos {\theta \over 2} =
\sqrt{N - M \over N}.
\end{equation}
The Grover iteration 
$\hat G = ( 2 | \psi_0 \rangle \langle \psi_0| - \hat I ) \hat O$,
where $\hat O$ denotes the oracle operator, 
performs the rotation by angle $\theta$ in the direction 
$| \alpha \rangle \to | \beta \rangle$
in the two-dimensional subspace spanned by 
$| \alpha \rangle$ and $| \beta \rangle$.
The state $| \psi_k \rangle$ 
after $k$ ($=0, 1, 2, \cdots$) iterations is therefore given by  \cite{NC}
\begin{equation}
| \psi_k \rangle =
\hat G^k | \psi_0 \rangle = 
\cos \left( {2k + 1 \over 2} \theta \right) | \alpha \rangle 
+ 
\sin \left( {2k + 1 \over 2} \theta \right) | \beta \rangle.
\label{eq:G^kpsiHT=}\end{equation}
Hence, by repeating the Grover iteration 
\begin{equation}
R \equiv \left\lceil {\arccos \sqrt{M/N} \over \theta} \right\rceil
\end{equation}
times, one can find a solution to the search problem with probability
$\gtrsim 1/2$.
Here, $\lceil a \rceil$ denotes 
the smallest integer among those larger than or equal to $a$. 
When $M \ll N$, in particular, 
we have $\theta \simeq 2 \sqrt{M/N}$ and 
\begin{equation}
R \simeq \left\lceil {\pi \over 4} \sqrt{N \over M}  \right\rceil.
\label{Rapprox}\end{equation}

\subsection{When $M=1$}\label{sec:M=1}

When $M=1$, 
Eqs.~(\ref{eq:alpha}) and (\ref{eq:beta}) reduce to
$| \alpha \rangle 
= | \psi_0 \rangle + O(1/\sqrt{N})
= | \psi_0 \rangle + O(1/2^{L/2})
$ 
and
$| \beta \rangle = | x_1 \rangle$,
respectively.
Since $| \psi_0 \rangle$ and $| x_1 \rangle$ are product states,
we find that $p=1$ for $| \psi_0 \rangle$, $| \alpha \rangle$
and $| \beta \rangle$.
For $| \psi_k \rangle$, on the other hand, 
Eq.~(\ref{eq:G^kpsiHT=}) gives the expectation value 
of the `$x$ component of the magnetization'
$\hat M_x = \sum_l \hat \sigma_x(l)$ as
\begin{eqnarray}
\langle \psi_k | \hat M_x | \psi_k \rangle 
&=&
\cos^2 \left( {2k + 1 \over 2} \theta \right) L + O(1).
\label{eq:pkMxpk}\end{eqnarray}
On the other hand, 
\begin{eqnarray}
\langle \psi_k | \hat M_x^2 | \psi_k \rangle 
&=&
\cos^2 \left( {2k + 1 \over 2} \theta \right) L^2 + O(L).
\label{eq:pkMx2pk}\end{eqnarray}
Equations (\ref{eq:pkMxpk}) and (\ref{eq:pkMx2pk}) yield
\begin{eqnarray}
\langle \psi_k | ( \Delta \hat M_x )^2 | \psi_k \rangle 
&=& 
{1 \over 4} \sin^2 \left( (2k + 1) \theta \right) L^2
+ O(L).
\label{eq:pkMx2pk-2}\end{eqnarray}
We thus find that 
$\langle \psi_k | ( \Delta \hat M_x )^2 | \psi_k \rangle = \mathcal{O}(L^2)$, 
i.e., $p=2$, 
if 
\begin{equation}
\sin^2 \left( (2k + 1) \theta \right) = O(1).
\label{eq:sin2=O(1)}\end{equation}
In a similar manner, we can also show that 
$\langle \psi_k | ( \Delta \hat M_z )^2 | \psi_k \rangle = \mathcal{O}(L^2)$,
where $\hat M_z = \sum_l \hat \sigma_z(l)$,
for these states.
[In order to show that $p=2$ for $| \psi_k \rangle$,
it is sufficient to find {\em one} additive observable 
which fluctuates macroscopically, 
as seen from the definition (\ref{def:p}).]

Since $\theta \simeq 2/\sqrt{N}$, 
condition (\ref{eq:sin2=O(1)}) is satisfied for {\em all} $k$ such that 
\begin{equation}
\delta \leq {4k+2 \over \sqrt{N}} \leq \pi - \delta,
\label{eq:cond-k}\end{equation}
where $\delta$ is an arbitrary small positive constant independent of $N$.
For example, all $k$'s such that 
$\lceil 0.1 R \rceil \leq k \leq \lceil 0.9 R \rceil$
satisfy this condition, 
according to Eq.~(\ref{Rapprox}).
We therefore conclude that most $| \psi_k \rangle$'s
in the Grover iteration processes  
are entangled macroscopically (i.e., $p=2$) when $M=1$, 
whereas $p=1$ for the initial and final states.

Note that the above result explicitly shows that 
$p$ is well defined for $| \psi_k \rangle$'s, 
although they are {\em not} strictly homogeneous in general.
We say this fact as `$| \psi_k \rangle$'s are {\em 
effectively homogeneous},'
as in Refs.~\cite{SM02,SS03,US04}.

\subsection{When $M=\mathcal{O}(1)$}\label{sec:M=O(1)}

When $M=\mathcal{O}(1)$ and $M \geq 2$, we have
$| \alpha \rangle = | \psi_0 \rangle + O(1/\sqrt{N})
= | \psi_0 \rangle + O(1/2^{L/2})
$ 
and
$| \beta \rangle = (| x_1 \rangle + \cdots + | x_M \rangle)/\sqrt{M}$.
Since $| \psi_0 \rangle$ is a product state, $p=1$ for $| \psi_0 \rangle$
and $| \alpha \rangle$.
However, unlike the case of $M=1$, 
$| \beta \rangle$ can have $p=2$.
For example, suppose that $M=2$ and the two solutions are 
$x_0 \equiv 1010 \cdots 10$ and $x_1 \equiv 0101 \cdots 01$.
Then, 
$| \beta \rangle = (| x_0 \rangle + | x_1 \rangle)/ \sqrt{2}$
is a state with $p=2$ because
$\langle \beta | ( \Delta \hat M_z^{\rm st} )^2 | \beta \rangle = \mathcal{O}(L^2)$,
where $\hat M_z^{\rm st} = \sum_l (-1)^l \hat \sigma_z(l)$.

For $| \psi_k \rangle$ with $k \geq 1$, on the other hand, 
Eqs.~(\ref{eq:pkMxpk})-(\ref{eq:pkMx2pk-2}) hold also in this case.
Hence, the discussion including 
Eqs.~(\ref{eq:sin2=O(1)}) and (\ref{eq:cond-k}) also holds.
Therefore, we conclude that 
{\em most $| \psi_k \rangle$'s
in the Grover iteration processes  
are entangled macroscopically, i.e., $p=2$, when $M=\mathcal{O}(1)$},
whereas $p=1$ for the initial state.
For the final state, $p$ depends on the nature of the solutions.

\subsection{When $M=\mathcal{O}(N)$}

The case of $M=\mathcal{O}(1)$ is most important and interesting
because condition (\ref{eq:faster}) is clearly satisfied.
In contrast, the cases of large $M$ such as 
$M=\mathcal{O}(\sqrt{N})$ and $M=\mathcal{O}(N)$ are uninteresting because
condition (\ref{eq:faster}) is not satisfied.
For completeness, however, we briefly discuss the case of 
$M=\mathcal{O}(N)$ as an example of such uninteresting cases.

When $M=\mathcal{O}(N)$, Grover's algorithm does not necessarily use 
macroscopically entangled states.
For example, 
suppose that $N$ is a multiple of $8$ and 
all multiples of $8$ less than $N$ are solutions.
Then, 
\begin{eqnarray}
| \alpha \rangle &=& 
\sqrt{8 \over 7N} \sum_{y=0}^{N/8-1} \sum_{z=1}^{7} | y z \rangle
=
\sqrt{1 \over 7}\sum_{z=1}^{7}| \rightarrow \cdots \rightarrow z \rangle,
\\
| \beta \rangle &=& 
\sqrt{8 \over N} \sum_{y=0}^{N/8-1} | y 000 \rangle
=
| \rightarrow \cdots \rightarrow 000 \rangle,
\end{eqnarray}
where $| \rightarrow \rangle = (|0 \rangle + |1 \rangle)/\sqrt{2}$.
Hence, all $| \psi_k \rangle$'s 
are product states except for the last three qubits.
Since entanglement of $\mathcal{O}(1)$ qubits cannot be 
the macroscopic entanglement, 
we find that $p=1$ for all $| \psi_k \rangle$'s.
This does not contradicts with the conjecture of Sec.~\ref{sec:conj}
because 
condition (\ref{eq:faster}) is not satisfied 
when $M=\mathcal{O}(N)$.

\subsection{Relevance of macroscopic entanglement to 
faster computation by Grover's algorithm}\label{sec:relevance-G}

From the results of Secs.~\ref{sec:M=1} and \ref{sec:M=O(1)}, 
we confirm the conjecture of Sec.~\ref{sec:conj} for 
Grover's quantum search algorithm, at least for the 
most important case of $M=\mathcal{O}(1)$.
On the other hand, 
as mentioned in Sec.~\ref{sec:conj}, 
the conjecture does {\em not} claim that 
{\em all} macroscopically entangled states appearing in 
the computation would be essential to fast quantum computation.
We now study this point.

When $M = \mathcal{O}(1)$, 
we have seen that 
most $| \psi_k \rangle$'s such that inequality (\ref{eq:cond-k}) is 
satisfied
are entangled macroscopically (i.e., $p=2$),
whereas 
the final state $| \psi_R \rangle$
may or may not be so 
depending on the number and natures of the solutions.
This fact suggests the following:
The macroscopic entanglement of most $| \psi_k \rangle$'s
should be 
relevant to the speedup by Grover's algorithm, 
whereas the macroscopic entanglement of the final state 
(and states close to the final state) 
should be irrelevant.
We argue that this is indeed the case.

The irrelevance of any entanglement of the final state is obvious.
In fact, to get a solution
one performs measurement 
on 
$| \psi_R \rangle \simeq 
| \beta \rangle = (| x_1 \rangle + \cdots + | x_M \rangle)/\sqrt{M}$.
Even if the quantum coherence among the states
$| x_1 \rangle, \cdots, | x_M \rangle$ are destroyed
by, say, external noises, the probability distribution of $x$, 
and thus the success probability of getting a solution, 
is almost unaffected because 
the measurement diagonalizes the computational basis,
which includes $| x_1 \rangle, \cdots, | x_M \rangle$.

In contrast, if the macroscopic entanglement of 
one of $| \psi_k \rangle$'s which satisfy inequality (\ref{eq:cond-k}) 
is destroyed, the success probability is significantly reduced.
For example, for 
$| \psi_{R/2} \rangle$ with $M=1$,
we have
\begin{equation}
| \psi_{R/2} \rangle 
\simeq
{1 \over \sqrt{2}} | \alpha \rangle 
+ 
{1 \over \sqrt{2}}  | \beta \rangle
\simeq
{1 \over \sqrt{2}} | \psi_0 \rangle 
+ 
{1 \over \sqrt{2}}  | x_1 \rangle,
\end{equation}
which is entangled macroscopically because
$| \psi_0 \rangle = |\rightarrow \rightarrow \cdots \rightarrow \rangle$
and $| x_1 \rangle$ are macroscopically distinct from each other.
If the quantum coherence between 
$| \psi_0 \rangle$ 
and $| x_1 \rangle$
are destroyed in $| \psi_{R/2} \rangle$, the state turns into the mixed state,
\begin{equation}
\hat \rho'_{R/2}
\simeq
{1 \over 2} | \psi_0 \rangle \langle \psi_0 |  
+ 
{1 \over 2} | x_1 \rangle \langle x_1 |.  
\end{equation}
If one applies the Grover iterations another $R/2$ times, this 
state evolves into
\begin{equation}
\hat \rho'_{R}
\simeq
{1 \over 2} | \psi_{R/2} \rangle \langle \psi_{R/2} |  
+ 
{1 \over 2} \hat G^{R/2} | x_1 \rangle \langle x_1 | \hat G^{\dagger R/2}.  
\end{equation}
When one performs measurement diagonalizing the computational basis 
on this degraded final state, 
the probability of getting the correct result $x=x_1$
is quite small because 
$\hat G^{R/2} | x_1 \rangle$ 
is much different from $| x_1 \rangle$.
Therefore, 
the macroscopic entanglement of $| \psi_{R/2} \rangle$, 
i.e., the quantum coherence between $| \psi_0 \rangle$ and $| x_1 \rangle$,
is crucial to faster computation.
In a similar manner, 
we can show that 
the macroscopic entanglement of most $| \psi_k \rangle$'s
is crucial to faster computation.

\section{Numerical results for Grover's quantum search algorithm}
\label{sec:Grover-numerical}

In the analysis of Sec.~\ref{sec:Grover-analytic}, 
we have considered entanglement of 
$| \psi_k \rangle =\hat G^k | \psi_0 \rangle$
($k = 0, 1, 2, \cdots, R$).
In actual quantum computations, 
the Grover iteration 
$\hat G$ 
is realized as a series of local 
and pair-wise operations \cite{NC}.
Hence, many intermediate states appear during the steps between 
$| \psi_k \rangle$ and $| \psi_{k+1} \rangle$.
Although the macroscopic entanglement of $| \psi_k \rangle$'s
is sufficient for confirming our conjecture,
we now study entanglement not only of 
$| \psi_k \rangle$'s but also of such intermediate states
in order to see details of evolution of entanglement.
For this purpose, we numerically simulate 
a quantum computer that performs 
Grover's quantum search algorithm.
As in the previous section, 
we consider the index register of $L$ qubits as a quantum computer, 
and thus Eq.~(\ref{eq:L=L0+O1}) is clearly satisfied.

%
%
%
%

\subsection{Formulation of simulation}

We simulate the case of $M=1$ 
because this case is most fundamental in the search problem.
The solution $x_1$ is chosen randomly. We have 
confirmed that this random choice of a solution
makes no significant differences on the 
results of the numerical simulations presented below.

As the initial state, a register ${\cal R}$ composed of $L$ qubits 
is set to be the following product state;
\begin{equation}
|\psi_{\rm init} \rangle = |0 \rangle.
\label{psi0_G}\end{equation}
Firstly, the Hadamard transformation is performed
by successive applications of the Hadamard gate on individual 
qubits in ${\cal R}$,  and the quantum state evolves into
\begin{equation}
|\psi_0 \rangle=
\frac{1}{\sqrt{2^{L}}}\sum_{x=0}^{2^{L}-1}|x\rangle.
\label{Hadamard}
\end{equation}
Then we apply the Grover iteration 
$\hat{G}= ( 2 | \psi_0 \rangle \langle \psi_0| - \hat I ) \hat O$,
which consists of two Hadamard transformations, 
an oracle operation $\hat{O}$, and a conditional phase shift $\hat{P}$, 
which works as
\begin{eqnarray}
\hat{P}|0\ket &=& |0\ket, \\
\hat{P} |x\ket &=& -|x\ket\ (x>0).
\end{eqnarray}
Each Hadamard transformation requires $L$ operations of 
the Hadamard gate. 
The oracle $\hat{O}$ is constructed depending on the function $f(\cdot)$,
and requires 
its own workspace qubits and computational steps.
However, since the oracle is not a proper part of the Grover's algorithm,
we simulate the operation of $\hat{O}$ as a one-step operation,
and its workspace is not included in the simulation.
The execution of the conditional phase shift $\hat{P}$ 
requires $\mathcal{O}(L)$ pairwise unitary operations.
For simplicity, we simulate $\hat{P}$ as 
a one-step operation. 
As a result, each Grover iteration is simulated by 
$2L+2$ steps of operations.
After the applications of the Grover iterations $R$ times,
the state $|\psi_0 \rangle$ evolves 
to 
\begin{equation}
\hat{G}^{R}|\psi_0 \rangle
=|\psi_R\rangle
\simeq |x_1\rangle.
\label{GRpsi0}\end{equation}
Finally, by performing a measurement on ${\cal R}$, 
one can obtain the solution
$x_1$ with a sufficiently high probability.
We do not simulate this measurement process.
The total computational steps $Q(L)$ in our simulation is thus
\begin{equation}
Q(L)=L+ (2L+2)R=\mathcal{O}(L\sqrt{2^{L}}).
\end{equation}

\subsection{Results of simulation}

\begin{figure}[htbp]
\begin{center}
\includegraphics[width=0.6\linewidth]{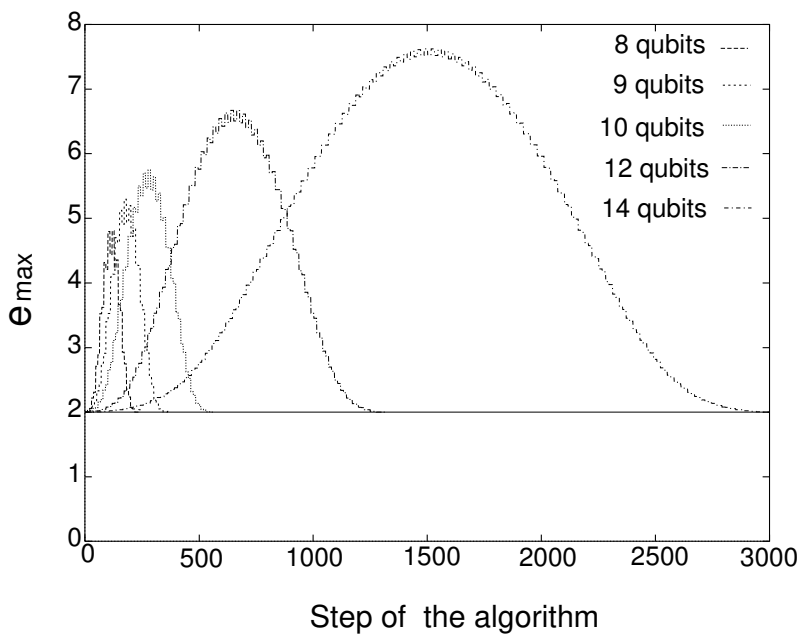}
\end{center}
\caption{
The maximum eigenvalue $e_{\rm max}$ of the VCM
of quantum states appearing in Grover's quantum search algorithm
for $L=8, 9, 10, 12, 14$
when $x_1 = 19, 388, 799, 1332, 9875$, respectively,
as functions of the step of the algorithm.
The horizontal line represents 
the value of $e_{\rm max}$ for product states, 
$e_{\rm max}=2.00$.}
\label{Grover9}
\end{figure}

\begin{figure}[htbp]
\begin{center}
\includegraphics[width=0.6\linewidth]{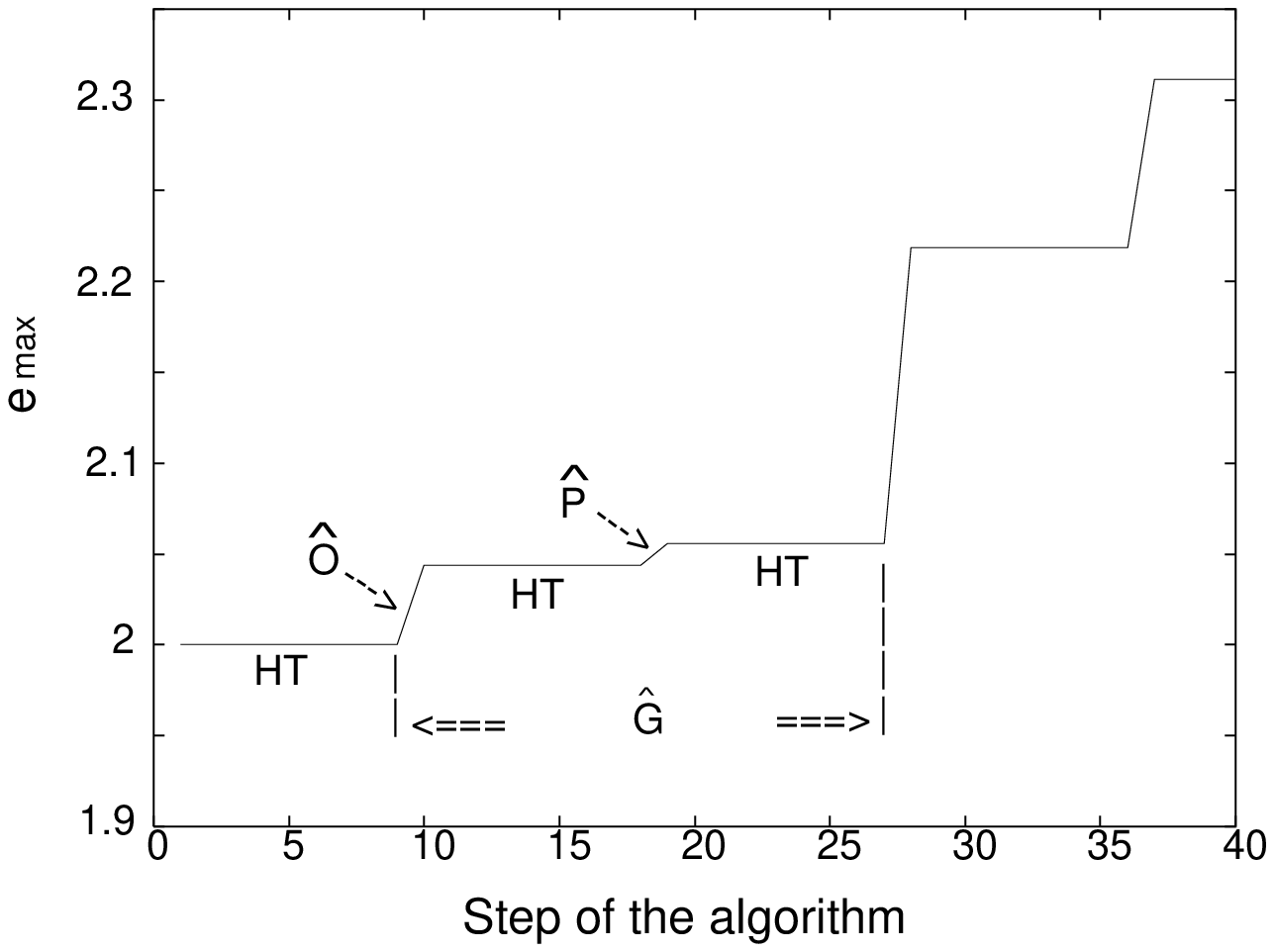}
\end{center}
\caption{
A magnification of Fig.~\ref{Grover9}, 
from the $1$st to $40$th steps for $L=8$. 
`HT' represents the Hadamard transformation, whereas
$\hat O$ and $\hat P$ represent
the oracle operation and conditional phase shift, respectively,
in a single Grover iteration  $\hat G$. 
}
\label{initial_G}
\end{figure}

\begin{figure}[htbp]
\begin{center}
\includegraphics[width=0.6\linewidth]{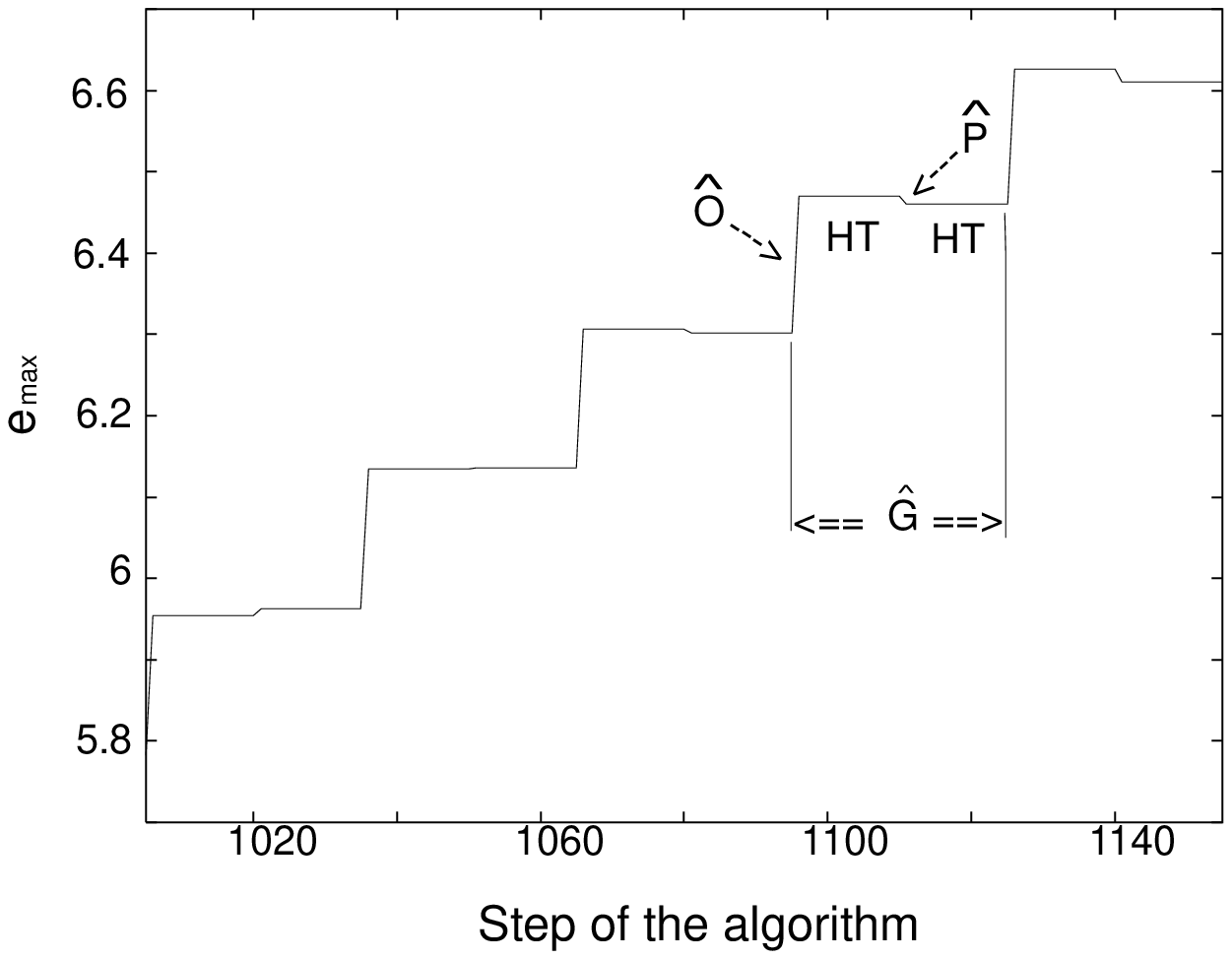}
\end{center}
\caption{
A magnification of Fig.~\ref{Grover9},
from the $1005$th to $1155$th steps for $L=14$. 
`HT' represents the Hadamard transformation, whereas
$\hat O$ and $\hat P$ represent
the oracle operation and conditional phase shift, respectively,
in a single Grover iteration  $\hat G$. 
}
\label{middle_G}
\end{figure}

Figure \ref{Grover9} plots $e_{\rm max}$ along the steps of 
Grover's algorithm, 
for $L=8, 9, 10, 12, 14$
when $x_1 = 19, 388, 799, 1332, 9875$, respectively.
Figure \ref{initial_G} is a magnification from the $1$st to $40$th steps
for $L=8$, 
whereas a magnification from the $1005$th to $1155$th steps for $L=14$
is shown in Fig.~\ref{middle_G}.


It is seen that 
$e_{\rm max}=2.00$ for all states from 
$|\psi_{\rm init} \rangle$ to $|\psi_0 \rangle$, 
i.e., during the initial Hadamard transformation, 
which is denoted as `HT' from the $1$st to $8$th steps
in Fig.~\ref{initial_G}.
This is because all these states are product states, 
for which we can easily show that $e_{\rm max}=2$ (Appendix \ref{appendix}). 
When the stage of Grover iterations begins, 
$e_{\rm max}$ grows gradually, as seen from 
Figs.~\ref{Grover9} and \ref{initial_G}.
In each Grover iteration, 
Figs.~\ref{initial_G} and \ref{middle_G} show that 
$e_{\rm max}$ changes when $\hat O$ is operated,
whereas it is kept constant during the subsequent Hadamard transformation.
Then, it changes again when $\hat P$ is operated,
whereas it is kept constant again 
during the subsequent Hadamard transformation.
As the Grover iterations are repeated, 
$e_{\rm max}$ continues to increase as a whole, 
until it takes the maximum value after about $R/2$ 
times applications of $\hat G$.
Further applications of $\hat G$ 
reduce $e_{\rm max}$, 
as seen from Fig.~\ref{Grover9},
toward $e_{\rm max} \simeq 2.00$ for $|\psi_R\rangle$, 
which is approximately a product state as seen from Eq.~(\ref{GRpsi0}).

\begin{figure}[htbp]
\begin{center}
\includegraphics[width=0.6\linewidth]{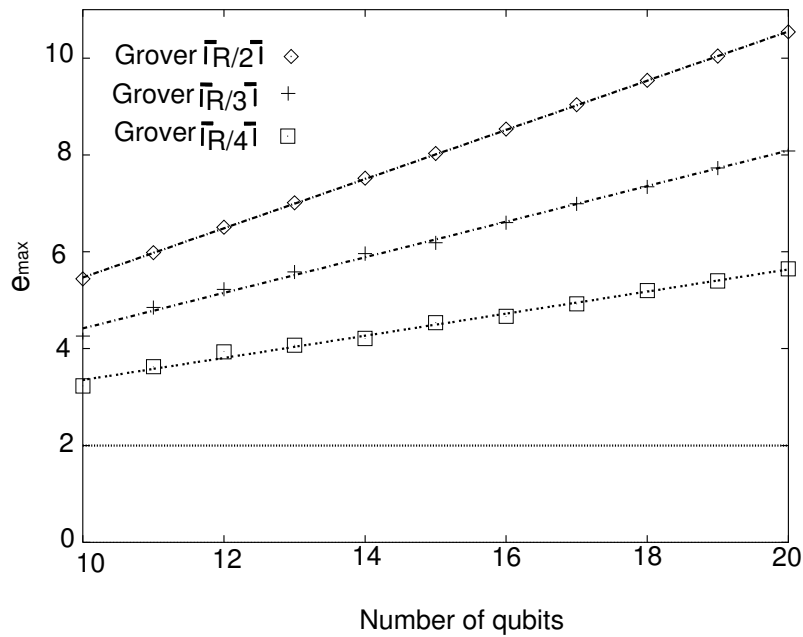}
\end{center}
\caption{
$e_{\rm max}$'s of $|\psi_{\lceil R/2 \rceil} \rangle$ (diamonds),
$|\psi_{\lceil R/3 \rceil} \rangle$  (crosses) 
and $|\psi_{\lceil R/4 \rceil} \rangle$ (squares),
as functions of $L$.
The dotted lines are the guides to the eyes,
whereas the horizontal line represents 
the value of $e_{\rm max}$ for product states, 
$e_{\rm max}=2.00$.
}
\label{maxfluG}
\end{figure}

To determine $p$, we note that 
$e_{\rm max}$'s are of the same order of magnitude 
for all states from $|\psi_k \rangle$ to $|\psi_{k+1} \rangle$
for each $L$,
as seen from Figs.~\ref{initial_G} and \ref{middle_G}.
Hence, they have the same value of $p$ as $|\psi_k \rangle$.
Therefore, from the result of the previous section, 
we conclude that 
all states from $|\psi_k \rangle$ to $|\psi_{k+1} \rangle$ have
$p=2$ for all $k$ which satisfies inequality (\ref{eq:cond-k}). 
As a demonstration, 
we plot $e_{\rm max}$'s of $|\psi_{\lceil R/2 \rceil} \rangle$,
$|\psi_{\lceil R/3 \rceil} \rangle$ and $|\psi_{\lceil R/4 \rceil} \rangle$
in Fig.~\ref{maxfluG} as functions of $L$.
Since $e_{\rm max}$'s are all proportional to $L$, 
we can confirm that $p=2$, i.e., these states are 
entangled macroscopically.
Note that Fig.~\ref{maxfluG} also demonstrates again 
that $p$ is well defined, although 
the states are not strictly homogeneous.

\subsection{Summary of analyses of Grover's quantum search algorithm}

The results of 
Secs.~\ref{sec:Grover-analytic} and \ref{sec:Grover-numerical}
for Grover's quantum search algorithm,
which is a representative algorithm for unstructured problems,
are summarized as follows.

When the number of solutions $M = \mathcal{O}(1)$, 
the search problem is hard in the sense that 
classical algorithms take $\mathcal{O}(N)$ steps.
This case is most important and interesting
because condition (\ref{eq:faster}) is clearly satisfied.
In this case, we have shown that 
most $| \psi_k \rangle$'s in Grover's algorithm 
are entangled macroscopically ($p=2$), 
and thus the conjecture of Sec.~\ref{sec:conj} is confirmed.

We have also found that 
the final state $| \psi_R \rangle$
may or may not be so depending on the number and natures of the solutions.
The conjecture does {\em not} claim that 
{\em all} macroscopically entangled states appearing 
the computation would be relevant to fast quantum computation.
We have shown that 
the macroscopic entanglement of most $| \psi_k \rangle$'s
is crucial to faster computation, whereas 
that of the final state is irrelevant.

The cases of large $M$ such as 
$M=\mathcal{O}(\sqrt{N})$ and $M=\mathcal{O}(N)$ are uninteresting 
in view of quantum computation because
condition (\ref{eq:faster}) is not satisfied.
For completeness, we briefly discuss the case of 
$M=\mathcal{O}(N)$ as an example of such uninteresting cases, 
and have shown that 
Grover's algorithm does not necessarily use 
macroscopically entangled states.

\section{Analysis of Shor's factoring algorithm}\label{sec:shor}

In this section, we study Shor's factoring 
algorithm \cite{Shor_1,EJ}, 
as a representative algorithm for structured problems \cite{NC}.

\subsection{Formulation of simulation}

Let ${\sf N}$ be a positive integer to be factored, 
${\sf x}$ a random number co-prime to ${\sf N}$ 
which satisfies $0<{\sf x}<{\sf N}$,
and $r$ the `order,' i.e., the least positive integer which satisfies 
${\sf x}^{r}\equiv 1$ ($\mathrm{mod}\ {\sf N}$). 
%
%
For a given pair of ${\sf N}$ and ${\sf x}$, one can  
find the value of $r$ efficiently using Shor's algorithm.
The first and second registers are denoted by 
${\cal R}_1$ and ${\cal R}_2$, respectively,
which have $L$ and $L'$ qubits.
We take \cite{L1=2L2}
\begin{equation}
\log_2 {\sf N} \leq L' < \log_2 {\sf N} +1, \quad L=2 L'. 
\label{L1L2}\end{equation}
We assume that 
this model satisfies Eq.~(\ref{eq:L=L0+O1}).
Note, however, that even if ${\cal R}_2$ could be replaced with a 
shorter register, the main result of this section would not change
because, as shown below,  only the qubits of ${\cal R}_1$ are entangled macroscopically.

The initial state
$|\psi_{\rm init}\ket$ is the product of two product states 
$| 0 \rangle^{(1)}$ and $| 1 \rangle^{(2)}$ of ${\cal R}_1$ and ${\cal R}_2$, 
respectively. 
We denote by $|\psi_{\rm HT}\ket$, $|\psi_{\rm ME}\ket$, and 
$|\psi_{\rm DFT}\ket$ 
the states just after the Hadamard transformation, 
modular exponentiation (ME), 
and discrete Fourier transformation (DFT), respectively.
For the DFT, we employ the quantum circuit of Sec.~V of Ref.~\cite{EJ},
which costs $L (L+1)/2$ steps.
The final states is $|\psi_{\rm DFT}\ket$.
 
As in Ref.~\cite{US04}, 
we represent the process 
$|\psi_{\rm HT}\rangle \to |\psi_{\rm ME}\rangle$
simply as the product of $L$ controlled unitary transformations, 
because ${\cal R}_1$ takes a major role \cite{US04}.
As a result, 
the states appearing during 
the modular exponentiation in our simulations 
correspond to $L$ representative states out of $\mathcal{O}(L^3)$ states.
As a result of this simplification, 
the total number $Q(L)$ of computational steps
in our simulation becomes
\begin{equation}
Q(L) = 2L + \frac{L (L+1)}{2},
\end{equation}
which is smaller than $\mathcal{O}(L^3)$ steps of a real computation.
To find the macroscopic entanglement, 
we shall evaluate $e_{\rm max}$ of the states that appear during the 
computation, as a function of 
$L_{\rm tot} = L+L' = 3L/2$.
[Since $L_{\rm tot} = \mathcal{O}(L)$, we will obtain 
the same results for $p$ if 
we evaluate $e_{\rm max}$ as a function of $L$.]

\subsection{Results of the simulation}

\begin{figure}[htbp]
\begin{center}
\includegraphics[width=0.6\linewidth]{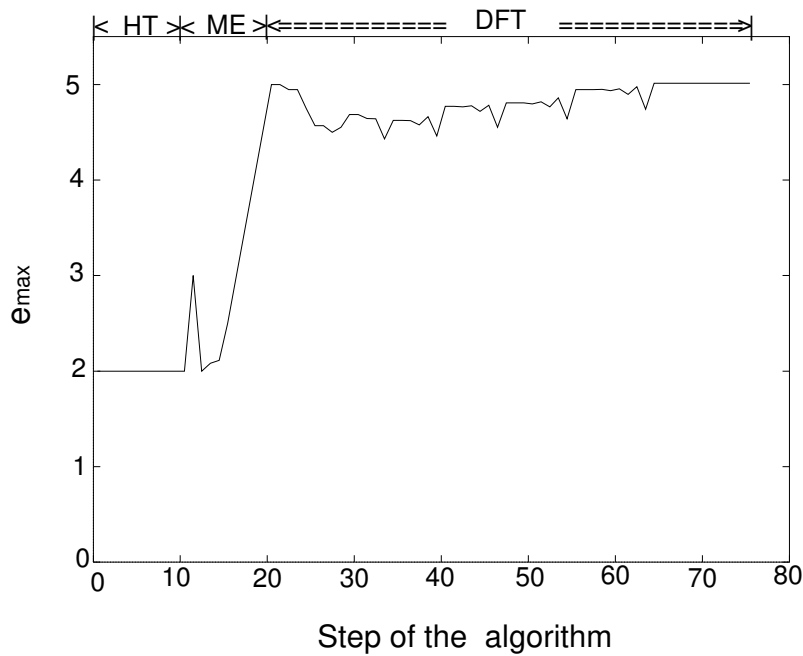}
\end{center}
\caption{
$e_{\rm max}$
of quantum states appearing in  Shor's factoring algorithm 
when 
${\sf N}=21,\ L_{\rm tot}=15,\ {\sf x}=2$, 
for which $r=6$,
as a function of the step of the algorithm. 
}
\label{N21y2r6}
\end{figure}

Figure \ref{N21y2r6} plots the maximum eigenvalue $e_{\rm max}$
of the VCM along the steps of the algorithm,
when 
${\sf N}=21,\ L_{\rm tot}=15,\ {\sf x}=2$, 
for which $r=6$. 
It is seen that 
$e_{\rm max}=2.00$ for all states from 
$|\psi_{\rm init} \rangle$ to 
the end of the Hadamard transformation. 
This is because all these states are product states,
for which $e_{\rm max}=2$ (Appendix \ref{appendix}). 
When the stage of the modular exponentiation begins, 
$e_{\rm max}$ grows, until it becomes $5.00$ for $|\psi_{\rm ME}\ket$.
Then, 
throughout the stage of the discrete Fourier transformation, 
$e_{\rm max}$ keeps large values,
slightly changing step by step.

\begin{figure}[htbp]
\begin{center}
\includegraphics[width=0.6\linewidth]{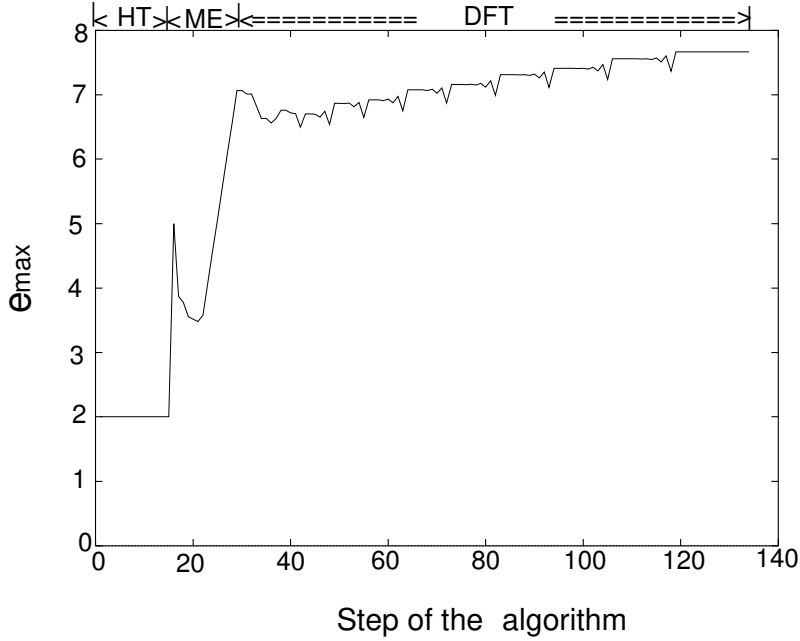}
\end{center}
\caption{
$e_{\rm max}$
of quantum states appearing in  Shor's factoring algorithm 
when 
${\sf N}=104,\ L_{\rm tot}=21,\ {\sf x}=55$, 
for which $r=6$,
as a function of the step of the algorithm. 
}
\label{N104y55r6}
\end{figure}

For other values of $({\sf N}, {\sf x})$,
and thus for other values of $L_{\rm tot}$ according to Eq.~(\ref{L1L2}),
we find that $e_{\rm max}$ behaves similarly
if $({\sf N}, {\sf x})$ gives $r=6$.
For example,  
Fig.\ \ref{N104y55r6} plots $e_{\rm max}$ when ${\sf N}=104,\ {\sf x}=55$,
for which $r=6$ again, and $L_{\rm tot}=21$. 
It is seen that $e_{\rm max}$ behaves similarly to the case of
Fig.~\ref{N21y2r6}.
Therefore, as in Ref.~\cite{US04}, we study the dependence of 
$e_{\rm max}$ on $L_{\rm tot}$ by varying $({\sf N}, {\sf x})$
{\em under the condition that it gives the same value of 
the order $r$}, $r=6$.
In this case, 
the states in Shor's algorithm become 
effectively homogeneous
for which $p$ is well defined,
as shown in Ref.~\cite{US04} and in the following.

\begin{figure}[htbp]
\begin{center}
\includegraphics[width=0.6\linewidth]{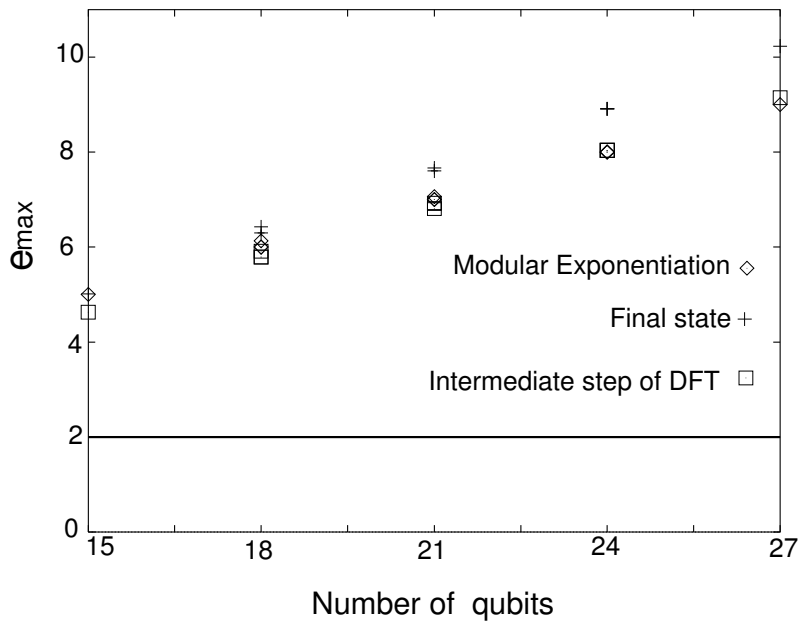}
\end{center}
\caption{
$e_{\rm max}$ of 
$|\psi_{\rm ME}\rangle$ (diamonds),
the state 
at the ${L \over 4} \left( {L \over 2} + 1 \right)$-th step 
of the DFT stage 
(squares), 
and the final state $|\psi_{\rm DFT}\rangle$ (crosses), for  $r=6$,
as functions of $L_{\rm tot} = L+L' = 3L/2$.
The horizontal line represents 
the value of $e_{\rm max}$ for product states, 
$e_{\rm max}=2.00$.
}
\label{maxfluS}
\end{figure}

Figure \ref{maxfluS} plots the $L_{\rm tot}$ dependence of $e_{\rm max}$
that is calculated in this way 
for three representative states including 
$|\psi_{\rm ME}\rangle$, 
the state at the ${L \over 4} \left( {L \over 2} + 1 \right)$-th step 
of the DFT stage 
(i.e, just after the pair-wise unitary transformations 
targeting $L/2$ qubits of ${\cal R}_1$ are finished),
and the final state $|\psi_{\rm DFT}\rangle$.
For each state, $e_{\rm max}$ increases linearly with increasing 
$L_{\rm tot}$.
Therefore, $p=2$ for these states.
In a similar manner, we find that 
all states after the modular exponentiation have $p=2$ for $r=6$, 
i.e., they are macroscopically entangled.
We have also obtained the same conclusion 
for other values of $r$ ranging from $2$ to $20$,
except when $r$ becomes an integral power of $2$
(i.e., $r = 2, 4, 8, 16$), for which we find that states with $p=2$ are not
necessarily used. 

Note that we already found in the previous work \cite{US04} that 
the {\em two} states, $|\psi_{\rm ME}\ket$ and $|\psi_{\rm DFT}\ket$,
are macroscopically entangled by studying {\em specific} additive operators.
In the present work, in contrast, 
we have proved the macroscopic entanglement of 
{\em all} states after the modular exponentiation,
by surveying {\em all} possible additive operators by the VCM method.
This demonstrates the power of the VCM method.

Another advantage of the VCM method is that 
one can identify the maximally-fluctuating additive operator(s)
from the eigenvector(s) corresponding to $e_{\rm max}$.
For example, $|\psi_{\rm ME}\ket$ for $r=6$ 
have two degenerate eigenvectors corresponding to $e_{\rm max}$,
from which 
we find that the following two operators 
are the maximally-fluctuating additive operators:
\begin{eqnarray}
\hat A_{\rm max}
&=&
\sqrt{\frac{3}{2}}\sum_{l = 2}^{L} (-1)^{l} \hat \sigma_y(l),
\label{A_1}
\\
\hat A_{\rm max}'
&=&
\sqrt{\frac{3}{2}}\sum_{l = 2}^{L} \hat \sigma_x(l).
\label{A_2}
\end{eqnarray}
For other values of $r$, other operators become 
the maximally-fluctuating additive operators.
In particular, the absence of operators of $l=1$ in 
Eqs.~(\ref{A_1}) and (\ref{A_2}) are accidental,
i.e., such operators are included in the sums 
of the maximally-fluctuating additive operators
for other values of $r$.
Note that 
$\hat A_{\rm max}$ and $\hat A_{\rm max}'$ do not contain 
operators of the second register $\mathcal{R}_2$.
This fact shows that the first register $\mathcal{R}_1$
are entangled macroscopically, whereas 
$\mathcal{R}_2$ is not, 
in consistency with the result of Ref.~\cite{US04}.

\subsection{Simulation with measurement after the modular exponentiation}

In the above simulations of Shor's algorithm,
measurement is not performed during the quantum computation.
In the original paper by Shor \cite{Shor_1}, on the other hand, measurement 
diagonalizing the computational basis is performed 
on ${\cal R}_2$ when the stage of the modular exponentiation is finished 
\cite{EJ}.
We also simulate this case.

\begin{figure}[htbp]
\begin{center}
\includegraphics[width=0.6\linewidth]{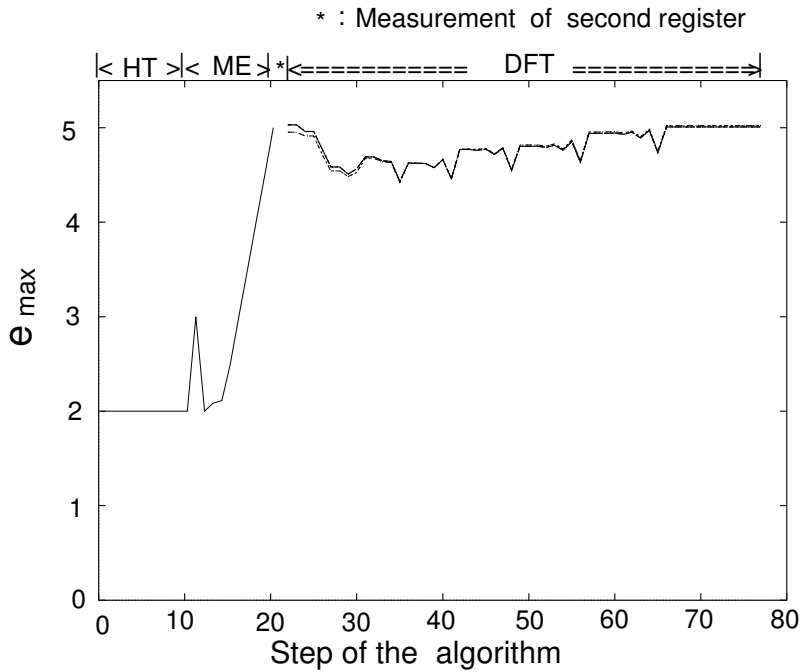}
\end{center}
\caption{
$e_{\rm max}$
of quantum states appearing in  Shor's factoring algorithm 
with measurement of the second register at the end of the 
modular exponentiation, 
when 
${\sf N}=21,\ L_{\rm tot}=15,\ {\sf x}=2$, 
for which $r=6$,
as a function of the step of the algorithm. 
After the measurement, denoted by the asterisk, 
two different profiles of $e_{\rm max}$
appear depending on the outcome of the measurement.
}
\label{N21y2r6_m}
\end{figure}

Figure \ref{N21y2r6_m} plots $e_{\rm max}$ 
under the same condition as Fig.~\ref{N21y2r6}
except that the 
measurement is performed at the end of the modular exponentiation.
By the measurement, 
the state of ${\cal R}_2$ is determined to be one of the states
 $|{\sf x}^a \mbox{ mod } {\sf N}\rangle^{(2)}$
where $a=1, 2, \cdots, r$.
Therefore, $e_{\rm max}$ can take $r$ different profiles
after the measurement.
Only two different profiles of $e_{\rm max}$
appear in Fig.~\ref{N21y2r6_m} because 
those for $a=1,2,3,4$ happen to be identical
and those for $a=5,6$ happen to be identical.
Both profiles of $e_{\rm max}$ are quite similar to 
the profile of Fig.\ \ref{N21y2r6}.
Therefore,
the measurement on ${\cal R}_2$ at the end of the modular exponentiation 
does not suppress the macroscopic entanglement.

\subsection{Relevance of macroscopic entanglement to 
faster computation by Shor's algorithm}\label{sec:relevance-S}

We have found that 
in Shor's factoring algorithm
{\em all} states after the modular exponentiation are macroscopically 
entangled
for almost all nontrivial values of $({\sf N},\ {\sf x})$, 
i.e., 
except for the special case where their values are such that the order 
$r$ becomes an integral power of $2$.
[This exceptional case will be discussed again in 
Sec.~\ref{sec:dc}.]
This again supports the conjecture of Sec.~\ref{sec:conj}.


However, as in the Grover's algorithm,
this does {\em not} mean that all of such states could be relevant to 
the faster computation.
In fact, we already showed in Ref.~\cite{US04} that the 
macroscopic entanglement of the final state $|\psi_{\rm DFT}\ket$ 
is irrelevant to the faster computation,
whereas that of $|\psi_{\rm ME}\ket$ is crucial.
This crucial macroscopic entanglement of $|\psi_{\rm ME}\ket$
does not disappear even if measurement is performed on the
second register just after the modular exponentiation is finished.
Therefore, as in the Grover's algorithm, 
macroscopic entanglement plays an essential role in 
Shor's factoring algorithm.

\section{discussions and conclusions}\label{sec:dc}

We have investigated macroscopic entanglement 
of quantum states in quantum computers that perform
Grover's quantum search algorithm
and Shor's factoring algorithm.
Here, we say a state is entangled macroscopically iff it has 
superposition of macroscopically states.
As a well-defined index of macroscopic entanglement, we have employed the 
index $p$ that was proposed and studied in Refs.~\cite{SM02,US04,MSS05,SS03}.
Using the method developed in Ref.~\cite{MSS05}, we have calculated 
this index as a function of the step of each algorithm.

For both algorithms, we have found that 
whether macroscopically entangled states are used or not depends on 
the numbers and properties of the solutions to 
the problem to be solved.
When the solutions are such that the problem becomes hard in the sense that 
classical algorithms take more than 
polynomial steps to find a solution, 
macroscopically entangled states are always used 
in Grover's algorithm and used 
almost always 
(i.e., except when the order $r$ is an integral power of $2$)
in Shor's algorithm.

Since Grover's and Shor's algorithms are representative ones for 
unstructured and structured problems, 
respectively, 
our results support strongly the conjecture 
(see Sec.~\ref{sec:conj}) 
that quantum computers should utilize 
macroscopically entangled states that is defined by $p$
in some stages of the computation
when they solve hard problems 
much faster than any classical algorithms.

Note that there are many different 
states with $p=2$ in many-qubit systems.
Which ones are used depends 
on the quantum algorithms and the inputs.
Note in particular that this conjecture 
does {\em not} claim that {\em all} states 
with $p=2$ would be useful in quantum computation.
Furthermore, 
among states 
with $p=2$ that do appear during quantum computation, 
some can be irrelevant to efficient computation,
as discussed in Secs.~\ref{sec:relevance-G} and \ref{sec:relevance-S} 
and in Ref.~\cite{US04}.

It is interesting to take the contraposition of the conjecture, i.e., 
{\em if macroscopically entangled states 
are not used throughout a quantum algorithm 
which satisfies Eq.~(\ref{eq:L=L0+O1}) 
then 
there must exist a classical algorithm that can solve the problem
as efficiently as the quantum algorithm, i.e., 
inequality (\ref{eq:faster}) is not satisfied}.
This may be used to explore {\em classical} algorithms with the help of 
knowledge about quantum algorithms.
For example, we have found that 
Shor's factoring algorithm does {\em not} necessarily use
macroscopically entangled states in the 
exceptional case where the order 
$r$ is an integral power of $2$. 
According to the above contraposition, this suggests that 
there would exist a classical algorithm that efficiently 
solve the problem in this special case.

 
\begin{acknowledgments}

We thank A. Hosoya for valuable discussions.

\end{acknowledgments}

\appendix

\section{Range of $p$}\label{appendix_range}

The range of $p$ is $1 \leq p \leq 2$.
For a system composed of L qubits, 
this can easily be shown as follows \cite{MSunpub}.
Let $e_1, e_2, \cdots$ be the eigenvalues of the
variance-covariance matrix $V$, Eq.~(\ref{VCM}), which is a
$3L \times 3L$ non-negative hermitian matrix.
Since
$
\sum_\alpha
\langle \psi(L) |(\Delta \hat{\sigma}_{\alpha}(l))^2
|\psi(L) \rangle
= \mathcal{O}(1)
$
for every $|\psi(L) \rangle$ and $l$, we find 
\begin{equation}
3L e_{\rm max}(L) \geq \sum_i e_i
= {\rm Tr} V
= \sum_l \sum_\alpha
\langle \psi(L) |(\Delta \hat{\sigma}_{\alpha}(l))^2
|\psi(L) \rangle
= \mathcal{O}(L).
\end{equation}
Therefore, according to Eq.~(\ref{emax=Lp-1}), 
$p \geq 1$.
On the other hand, $p \leq 2$ because
\begin{equation}
\left| \langle \psi(L) | 
\Delta \hat A^\dagger \Delta \hat A
| \psi(L) \rangle \right| 
\leq
\sum_{l, l'} \sum_{\alpha, \alpha'} 
\left| 
\langle \psi(L) | 
c^*_{l \alpha} \Delta \hat{\sigma}_{\alpha}(l) 
c_{l' \alpha'} \Delta \hat{\sigma}_{\alpha'}(l')
| \psi(L) \rangle \right|
\leq
\mathcal{O}(L^2).
\end{equation}
These arguments can be easily generalized to more general systems.

\section{Maximum eigenvalue of the VCM for product states}\label{appendix}

In this appendix, we show that $e_{\rm max} = 2$ for
a pure state $|\psi\rangle$ if it is a product state, 
\begin{equation}
|\psi\rangle
=
\bigotimes_{l=1}^L |\phi_l \rangle_l,
\end{equation}
where $|\phi_l \rangle_l$ denotes a state of 
the qubit at site $l$
($=1, 2, \cdots, L$).
The VCM of such a state becomes a block-diagonal matrix
\begin{equation}
\left(
\begin{array}{ccccc}
\bm{V}_1 & 0 & \cdots & \cdots & 0\\
0 & \bm{V}_2 & 0 & \cdots & \cdots\\
\vdots & 0 & \bm{V}_3 & 0 &\cdots \\
\vdots & \vdots & & \cdots & \cdots \\
0 & \cdots & \cdots & 0 & \bm{V}_L 
\end{array}
\right),
\end{equation}
where $\bm{V}_l$ is a $3 \times 3$ matrix 
whose $\alpha \beta$ element ($\alpha, \beta = x, y, z$) is given by
\begin{eqnarray}
(V_l)_{\alpha \beta}
&=&
\langle \psi|\hat{\sigma}_\alpha(l)\hat{\sigma}_\beta(l)|\psi\rangle
-\langle \psi|\hat{\sigma}_\alpha(l)|\psi\rangle
\langle\psi|\hat{\sigma}_\beta(l)|\psi\rangle
\\
&=&
\langle \phi_l|\hat{\sigma}_\alpha(l)\hat{\sigma}_\beta(l)|\phi_l\rangle
-\langle \phi_l|\hat{\sigma}_\alpha(l)|\phi_l\rangle\langle\phi_l|\hat{\sigma}_\beta(l)|\phi_l\rangle.
\end{eqnarray}
Therefore, $e_{\rm max}$ is given by 
the maximum one among the maximum eigenvalues of 
$\bm{V}_l$'s.
By a unitary transformation of this $3 \times 3$ matrix
such that $|\phi_l\rangle$ becomes an eigenstate 
of the transformed $\hat \sigma_z(l)$, 
we can transform $\bm{V}_l$ into 
\begin{equation}
\bm{V}_l=
\left(
\begin{array}{ccc}
1 & i & 0\\
-i & 1 & 0 \\
0 & 0 & 0   
\end{array}
\right).
\end{equation}
Since the maximum eigenvalue of this matrix is $2$, 
we find that $e_{\rm max} = 2$.


\begin{thebibliography}{99}
\bibitem{custom}
Although the macroscopic entanglement 
is defined for a family of similar states, 
it is customary to talk about the macroscopic entanglement 
of a state of a particular size \cite{SM02,MSS05,SS03}. 
This is the same custom as the computational 
complexity: Although it is defined for a family of 
various sizes of the input, it is customary to talk about
the computational complexity for a particular size.
For example, one often says that factoring a number of $1024$ bits 
is hard.
\bibitem{O}
In this paper, we use two symbols 
$\mathcal{O}$ and $O$ to represent 
asymptotic behavior of a function $f(L)$
as $L \to \infty$:
$f(L)=\mathcal{O}(L^n)$ if $\lim_{L\to\infty}f(L)/L^n=$ constant $\neq 0$, 
whereas 
$f(L)=O(L^n)$ if $\lim_{L\to\infty}f(L)/L^n$ is finite.
\bibitem{CPandp}
In Ref.~\cite{SM02}, pure states with $p=2$ are called 
anomalously-fluctuating states.
\bibitem{L1=2L2}
Equation (16) of Ref.~\cite{US04} has a typo:
Actual calculations were performed by taking 
$L$ and $L'$ as given by Eqs.~(\ref{L1L2}) of the present paper.
Anyway, such small difference of $O(1)$ does not matter
to discuss $p$.
\end{thebibliography}
\end{document}